 \documentstyle[12pt]{article}
\global\arraycolsep=2pt

\newcommand{\k}{\vec{k}_{\perp}^2}
\newcommand{\be}{\begin{eqnarray}}
\newcommand{\ee}{\end{eqnarray}}
\newcommand{\la}{\langle}
\newcommand{\ra}{\rangle}

\newcommand{\Dm}{\vec{iD}_{\mu} }
\begin{document}

\begin{titlepage}
\begin{flushright}
SMU-HEP-94-19\\
hep-ph/9410228\\
October 1994
\end{flushright}

\vspace{0.3cm}
\begin{center}
\Large\bf  
Quark Model, \\ Large Order Behavior\\ 
and Nonperturbative Wave Functions  
  in QCD.\\
   \end{center}

\vspace {0.3cm}

 \begin{center} {\bf Ariel R. Zhitnitsky\footnote{
On leave of absence from Budker Institute of Nuclear Physics,\\ 
Novosibirsk,630090,Russia.\\
e-mail addresses:arz@mail.physics.smu.edu, ariel@sscvx1.ssc.gov}}
 \end{center}
 
\begin{center}
{\it Physics Department, SMU , Dallas, Texas, 75275-0175}

\end{center}
\begin{abstract}
 
We discuss a
 few, apparently different
(but actually, tightly related) problems: \\
1. The relation between QCD and valence quark model,\\
2. The evaluation of the nonlocal condensate
 $ \la \bar{q}(x)q(0)\ra $,
  its relation to heavy-light $\bar{q}Q$ quark system
and to constituent quark mass, \\
3. The asymptotic behavior of  the  nonperturbative
pion wave function $\psi(\k, x)$ at $x\rightarrow 0,~1, \k
\rightarrow \infty$  and  \\
4. The large order behavior of perturbative series. 
 
The analysis is
based on such general methods as    dispersion relations,
duality and  PCAC.
 We use  the steepest descent  
method (also known as semiclassical, or instanton calculus),
introduced by Lipatov
  to calculate  the $n-$th  moment of the
$\psi(\k, x)$ with result $\la\vec{k}_{\perp}^{2n}\ra \sim n!$.
This information  
 fixes of the asymptotic behavior of $wf$ at large $\k$.
   This behavior 
 is turned    out to be Gaussian 
    commonly used in the phenomenological analyses.
 The same method determines the asymptotic behavior
of the 
mixed local vacuum condensates 
$\la\bar{q}G_{\mu\nu}^nq\ra\sim n!$ at large $n$ as well as the 
nonlocal vacuum condensate
$ \la \bar{q}(x)q(0)\ra $
which   naturally arises in the description of the
heavy-light $\bar{q}Q$ quark system.

The relation 
between nonlocal condensate and constituent quark mass is also discussed.

\end{abstract}
\end{titlepage} 
\vskip 0.3cm
\noindent
{\bf 1. Introduction}

 The problem of bound states in the
relativistic quantum field theory with
large coupling constant is an extremely difficult
problem. To  understand the structure of the
bound state is a very ambitious goal which
assumes the 
solution of a whole spectrum of tightly connected problems, such as 
confinement,
chiral symmetry breaking phenomenon,   and many others which
are greatly important in the low energy region. 

A less ambitious purpose is the  study of
 the hadron wave function ($wf$) with a minimal number 
of constituents
( we consider such an analysis as the possible 
  way of  understanding  the valence quark model in   terms of QCD.).
As is known such a function gives the  parametrically leading contributions 
to hard exclusive processes.  The corresponding 
wave functions within QCD  have  been introduced to the theory
 in the late seventies
and  early eighties  \cite{Brod}   to describe the
exclusive processes. We refer to the review papers 
\cite{Cher},\cite{Brod1}, \cite{Cher1}
 on this subject for the  details.
  
The main idea of the approach \cite{Brod}-\cite{Cher1}
 is the separation of the large and small
distance physics. At small distances we can use the standard
perturbative expansion due to the asymptotic freedom and 
smallness of the coupling constant. All nontrivial,
large distance physics is hidden into the  
nonperturbative wave function ($wf$)  
and can not be found 
by perturbative technique, but rather it should be extracted
from  elsewhere.   The most powerful analytical 
nonperturbative method
for such   problems, I believe, is the 
QCD sum rules \cite{Shif1},\cite{Shif2}.

The first application of QCD sum rules to the analysis
of nonperturbative $wf$ was considered more than decade ago
\cite{Cher2}.  The information   extracted
for the few lowest
moments, unambiguously shows the asymmetric form of the
distribution amplitudes.  Since then this subject
  is a very controversial
issue   \cite{Mikh}-\cite{Cotanch} and we are not going
to comment  these quite opposite  points  
in the present note. However we would like to 
make a comment that a finite number of moments is not
sufficient to completely determine the $wf$. The behavior
of the asymptotically distant terms plays  a crucial role
in the reconstruction of the $wf$.

$\bullet$ The main goals of the paper are:

 The   analysis of  
the asymptotically distant
terms at $n\rightarrow\infty$ in the longitudinal $\la (2x-1)^n\ra $
 and transverse
$\la \vec{k}_{\perp}^{2n} \ra  $
directions. We are going to use  very general methods
in this analysis, like dispersion relations and duality.
Thus, the obtained results should be considered
as the direct consequences of QCD.

Besides that, we use the same methods  to analyse the so-called
nonlocal condensate (NLC)
$\la\bar{q}(x)P\exp(ig\int_0^xA_{\mu}dx_{\mu})q(0)\ra$
which is the fundamental object of the QCD-theory.
It  naturally appears in the analysis of the heavy-light quark
system.
 
We make contact between  valence quark model 
and light cone $wf$ defined exclusively
in QCD terms. Finally, we define and calculate
a constituent mass of the quark in terms of NLC,
i.e. in terms of the object which is defined itself in terms of QCD.

 Let me emphasize from the 
very beginning  that 
 the ideology and methods (unitarity,
dispersion relations, duality) we use are motivated
by QCD sum rules.  However,  we do not use the  QCD sum rules
in the common sense: we do not fit them  
  to extract any information about lowest resonance 
(as  people usually  do in this approach), we do not
use any numerical approximation or implicit assumption about
higher states. Instead, we concentrate on analysis of the
appropriate correlation functions themselves   to extract
 the most general information.
 
The paper is organized as follows.
In the next section we define the nonperturbative $wf$ through its moments.
 
Section 3 is 
devoted to the analysis of the large order behavior. 
 We use the steepest descent  
method (also known as semiclassical, or instanton calculus)
introduced by Lipatov,
  to calculate  the $n-$th order  of perturbative series.
After that the dispersion relations
unambiguously fix the asymptotic behavior of $wf$ at large $\k$.
 
In   section 4 we apply the same method for   consideration
of the different object - the   NLC,
which originally was introduced for analysing
the heavy-light quark system.  
  \vskip .3cm
\noindent
{\bf 2. The Definitions. The constituent quark model and QCD.}
  \vskip .3cm

We define the pion
axial wave function 
in the following gauge-invariant way:
\be
\label{d}
if_{\pi}q_{\mu}\phi_A (zq,z^2)=
\la 0|\bar{d}(z)\gamma_{\mu}\gamma_5 
e^{ig\int_{-z}^z A_{\mu}dz_{\mu}} u(-z)|\pi(q)\ra \\
\nonumber 
=\sum_n \frac{i^n}{n!}\la 0|\bar{d}(0)\gamma_{\mu}\gamma_5
(iz_{\nu}\stackrel{\leftrightarrow}{D_{\nu}})^n u(0)|\pi(q)\ra ,
\ee
where 
$\stackrel{\leftrightarrow}{D_{\nu}}\equiv
\stackrel{\rightarrow}{D_{\nu}}-\stackrel{\leftarrow}{D_{\nu}}$ and 
$\Dm=i\vec{\partial_{\mu}}+gA_{\mu}^a\frac{\lambda^a}{2}$ is the
covariant derivative. 
From its definition is clear that the set of different
$\pi$ meson matrix elements defines the nonperturbative wave
function.  

The most important part (at  asymptotically high $q^2$) is
the one related to the longitudinal distribution. 
In this case $z^2\simeq 0$ and the $wf$
depends only on one $zq$- variable. The corresponding
Fourier transformed
wave function will be denoted as $\phi(\xi)$ 
and its 
  $n-$th moment is given by the
following local matrix element: 
\be
\label{1}
\la 0|\bar{d}\gamma_{\nu}
\gamma_5(i\stackrel{\leftrightarrow}{D_{\mu}}
z_{\mu})^{n} u|\pi(q)\ra=if_{\pi}q_{\nu}
(zq)^{n} \la \xi^{n}\ra=if_{\pi}q_{\nu}(zq)^{n}
\int^1_{-1}d\xi\xi^{n}\phi(\xi)
\ee
\be
\label{}
-q^2\rightarrow\infty,~~zq\sim 1~~
\xi=x_1-x_2,~~ x_1+x_2=1,~~z^2=0.   \nonumber
\ee
 Therefore, if we knew all matrix elements (\ref{1})
(which are perfectly well-defined) we could restore the whole
$wf$. The QCD sum rules approach allows one to find the 
magnitudes only the few first moments \cite{Cher2}. 
As is known, this information is  not enough to reconstruct
the $wf$; the parametric behavior at $\xi\rightarrow\pm 1$
is the crucial issue in this reconstruction.
To extract the corresponding information,
let us consider the following correlation function:
\be
\label{2}
i  \int dx e^{iqx}\la 0|T J_{n}^{\|}(x),J_0(0) |0\ra=
(zq)^{n+2}I_{n}(q^2),~~
 J_{n}^{\|}=\bar{d}\gamma_{\nu}z_{\nu}
\gamma_5(i\stackrel{\leftrightarrow}{D_{\mu}}z_{\mu})^{n} u 
 \ee
and calculate its asymptotic behavior at large $q^2$.
The result can be presented in the form of the dispersion integral,
whose spectral density is determined by the pure perturbative one-loop
diagram (Fig.1):
\be
\label{3}
\frac{1}{\pi}\int_0^{\infty} ds\frac{Im I_n^{pert}(s)}{s-q^2},~~
 Im I_{n}(s)^{pert}=\frac{3}{4\pi(n+1)(n+3)}.
 \ee
 Our main assumption  corresponds
to the very likely case that the $\pi$ meson gives a
nonzero contribution to the dispersion integral for arbitrary
$n$ and, in particular, for $n\rightarrow\infty$.
Formally, it can be written in the following way
 \be
\label{3a}
 \frac{1}{\pi}\int_0^{S_{\pi}^n} ds Im I(s)^{pert}_{n}=
\frac{1}{\pi}\int_0^{\infty} ds Im I(s)^{\pi}_{n},
 \ee
We assume that ${S_{\pi}^n}$ in this relation is some finite, nonzero number 
(we can say that `` the pion
 fills"  a finite duality interval).
  Our assumption means
  that there are no special cancelations
and $\pi$ meson contribution to the
dispersion integral is not zero, i.e. $S_{\pi}^n(\|)\neq 0$,
where we specified the notation for the longitudinal distribution.
In this case at $q^2\rightarrow\infty$ our assumption (\ref{3a})
leads to the following relation:
\be
\label{4}
  f_{\pi}^2\la \xi^n\ra (n\rightarrow\infty)
\rightarrow\frac{3S_{\pi}^{\infty}(\|)}{4\pi^2n^2}
 \ee
It  unambiguously  implies the  following behavior at
the end-point region \cite{Cher}: 
 \be
\label{5}
 \la\xi^n\ra=\int_{-1}^1d\xi \xi^n\phi(\xi)\sim 1/n^2,~~~~~  
   \phi(\xi\rightarrow
\pm 1)\rightarrow (1-\xi^2).
\ee
Let us note, that the same arguments for the 
pseudoscalar $wf$ imply its constant behavior at 
$\xi\rightarrow\pm 1$. 

Now we want to repeat these arguments for the analysis of the 
transverse distribution. To do so, let us define
the mean values of the transverse quark distribution
by the following matrix elements:
\be
\label{6}
\la 0|\bar{d}\gamma_{\nu}
\gamma_5 (\stackrel{\rightarrow}{iD_{\mu}}
 t_{\mu})^{2n} u|\pi(q)\ra=if_{\pi}q_{\nu}
 (-t^2)^n\frac{(2n-1)!!}{(2n)!!}\la \vec{k}_{\perp}^{2n} \ra.
\ee
where $\stackrel{\rightarrow}{D_{\nu}}$ is the
covariant derivative,
acting on the one quark  and  
 transverse vector $t_{\mu}=(0,\vec{t},0)$ is perpendicular
 to the hadron momentum $q_{\mu}=(q_0,0_{\perp},q_z)$. 
The factor $\frac{(2n-1)!!}{(2n)!!}$ is introduced   to
(\ref{6}) to take into account  
  the integration over $\phi$ angle in the transverse plane:
$\int d\phi (\cos\phi)^{2n}/ \int d\phi= {(2n-1)!!}/{(2n)!!}$.
 
We interpret the $\la \vec{k}_{\perp}^{2} \ra$ in this equation
as  a mean value of the quark perpendicular momentum. Of course it
is different from the naive, gauge dependent  definition like
$\la 0|\bar{d}\gamma_{\nu}
\gamma_5 \partial_{\perp}^2 u|\pi(q)\ra $,
because the physical transverse gluon is participant 
of this definition.
However, the expression (\ref{6}) is the only possible 
way to define the  $\k$
in the gauge theory like QCD.
 We believe that such definition is the useful generalization
of the transverse momentum conception for the interactive quark
system.

The first few  transverse moments 
defined above, can be calculated   using some special 
  QCD sum rules \cite{Zhit3}. 
 To find the behavior $\la \vec{k}_{\perp}^{2n} \ra$ at large $n$
  we can repeat our previous 
duality arguments with the following result\footnote{ Here 
and in what follows we ignore any mild (nonfactorial) $n$-dependence.
As we   discuss later, we will be able to
find a factorial dependence, but not a prefactorial behavior. Thus,
they will be consistently omitted.}:
\be
\label{8}
f_{\pi}^2\la \vec{k}_{\perp}^{2n} \ra \frac{(2n-1)!!}{(2n)!!}\sim
\frac{3S_{\pi}^{n+1}(\perp)(2n+2)!!}{8 \pi^2(2n+3)!!}\Rightarrow 
 f_{\pi}^2\la \vec{k}_{\perp}^{2n} \ra \sim S_{\pi}^{n+1}(\perp). 
\ee
Here we introduced the   notation  $S_{\pi}(\perp)$
for the $\pi$ meson interval duality  
(do  not confuse  with 
corresponding notation for the longitudinal
distribution $S_{\pi}(\|)$).
 
The nice feature of   (\ref{8})
is its finiteness  for arbitrary $n$. It means 
that the higher  moments 
$$ 
\la \vec{k}_{\perp}^{2n} \ra =
\int d\k d\xi
\vec{k}_{\perp}^{2n}\psi(\k, \xi ) $$ 
{\bf do exist}.
In this formula we introduced the nonperturbative $wf$
$\psi(\k, \xi )$, normalized to one. 
Its moments are determined by the local
matrix elements (\ref{6}). The relations to Brodsky and Lepage
notations $\Psi_{BL}(x_1,\vec{k}_{\perp})$
\cite{Brod1} and to longitudinal distribution
amplitude $\phi(\xi)$ introduced earlier, look as follow:
\be
\label{9}
\Psi_{BL}(x_1,\vec{k}_{\perp})=\frac{f_{\pi}16\pi^2
}{\sqrt{6}}\psi(\xi,\vec{k}_{\perp}),~
\int d\k \psi(\k, \xi )
= \phi(\xi),~\int_{-1}^1 d\xi\phi(\xi)=1 
\ee
where $f_{\pi}=133 MeV$.
The  existence of the arbitrary high moments  
$\la \vec{k}_{\perp}^{2n} \ra$ means that the nonperturbative
$wf$, defined above, falls off at large transverse momentum $\k$
faster than any power function (exponential?).

Another feature of the formula (\ref{8}) 
is apparently the bad news,
because it corresponds
to a  nonphysical $\delta (\k-S_{\pi}(\perp))$-shape distribution.
It is absolutely clear that the physical interactions spread out 
this behavior and  will change this result.\footnote{
The situation here in some sense resembles the deep inelastic
scattering. As is known the asymptotic form of the
nucleon structure function at $q^2\rightarrow\infty$ is
$F_2(x,q^2\rightarrow\infty)\sim\delta(x)$. However,
$F_2$ differs strongly from the delta function 
and goes to this limit only at asymptotically high $q^2$
when the terms like $1/\ln q^2$ are  neglected.}
Indeed, as will be shown in the next section, the higher loop
corrections give a desired factorial behavior on $n$.
 
To anticipate the events we would  like to formulate here
the obtained result. Higher loop corrections will change the
behavior (\ref{8}) in the following way:
\be
\label{10}
 S_{\pi}^{n+1}(\perp) \Rightarrow
 S_{\pi}^{n+1}(\perp)  \{1+...+ c_n(\frac{\alpha_s}{\pi})^n
 n  !\},~~c_n\sim 1.
\ee
For the large enough $n$   the perturbative corrections
will play the dominant role and  they actually
define the large $\k$ behavior of the nonperturbative $wf$,
which   turns out to be Gaussian.
 The relation (\ref{8}) converts this information into the 
fixing of the asymptotic behavior of $wf$ at large $\k$:
\be
\label{!}
 \la \vec{k}_{\perp}^{2n} \ra =
\int d\k d\xi
\vec{k}_{\perp}^{2n}\psi(\k, \xi ) \sim  n ! 
~~n\rightarrow\infty \\  \nonumber
\Longrightarrow
\psi(\k\rightarrow\infty, x)\sim \exp(-\frac{\k}{x(1-x)} ),
 \ee
where we took into account the previous result \cite{Zhit3} that 
$\frac{\k}{x(1-x)}$ is the only
allowed combination the $wf$ can depend on.

Few comments are in order. 
 We {\bf identify} the moments (\ref{6}) defined in QCD with
the ones    defined  in the constituent quark model.
We believe that such an identification is the useful working conjecture,
which might be helpful in understanding   the connection  
between QCD and the valence quark model.

 Now let me recall  few results (from the constituent quark model) which 
support this conjecture. 
It is well known \cite {Isgur1} that the equal- time
 wave functions 
\be
\label{qm}
\psi_{CM}(\vec{q}^2)\sim \exp (-\vec{q}^2)
\ee
of the harmonic oscillator in the rest frame
give a very reasonable description of
static meson properties.
Together with Brodsky-Huang-Lepage prescription \cite{BHL}
connecting  the equal -time   and the 
light-cone wave functions
by identification
$$ \vec{q}^2\leftrightarrow\frac{\k}{4x(1-x)},
~~\psi_{CM}(\vec{q}^2)\leftrightarrow\psi_{LC}(\frac{\k}{4x(1-x)}),$$
one can reproduce the 
Gaussian behavior (\ref{!}) found exclusively from QCD\footnote{More
 exactly, in this identification the combination
with constituent effective
quark mass $\k+m^2$ appeares. However, because we are interested in
the asymptotic $\k$ behavior, we skipped these terms.}.
It means, first of all, that our identification 
of the moments (\ref{6}) defined in  QCD
with the ones defined in quark model,
is the reasonable conjecture
\footnote{The same method can be applied for the analysis
of the asymptotical behavior of the nucleon $wf$ which
in obvious notations takes the form:
$$\psi_{nucleon}(\vec{k}_{\perp i}^2 
\rightarrow\infty, x_i)\sim \exp(-\sum\frac{ \vec{k}_{\perp i}^2}{x_i})$$.}.
 
     \vskip .3cm
\noindent
{\bf 3.Perturbative series at large orders.}
  \vskip .3cm

This section is the  technical one. Its purpose is the 
explanation of the formula (\ref{10}) which constraints
the  asymptotics of the $wf$ in the transverse  direction.
 
The main idea, as before, is analysis of the appropriate correlation
functions 
at large $q^2$. We estimate the perturbative corrections 
at large $n$   to find  the dependence on 
external number $n$.
 Our main assumption  corresponds
to the very likely case that the $\pi$ meson gives a
nonzero contribution to the dispersion integral for arbitrary
$n$.
Formally, it can be written in  
  terms of dispersion relation (\ref{3a}).
We assume that ${S_{\pi}}$ in this relation is some finite, nonzero number. 
 As soon as this assumption has been
made, the standard calculation of the $n$ dependence
of  correlation function   
 convert this information into the knowledge about
$n$ dependence for the $\pi$ meson matrix elements (\ref{1},\ref{6}).
This information is required for the fixing  of the $wf$
asymptotic behavior.

Before going on, let me pause here  to recall
Lipatov's method \cite{Lipatov} 
of the calculation of  large order behavior
 in a field theory (for recent review
of this technique, see \cite{JZJ}). For great simplification we would like
to demonstrate the main idea of the method in a scalar field theory.
The generalization for the nonabelian gauge theories is straightforward.

The main object of our analysis  is the $n$-point correlation function,
$Z^{(n)}(g)$,
where $n$ is the number of external fields (external currents)
and $g$ is the coupling constant in  a scalar field theory.
For $g$ small, $Z^{n}(g)$ is given by an asymptotic expansion:
\be
\label{14}
 Z^{(n)}(g)=\sum_{k} Z_{k}^{(n)} g^k
\ee
We would like to calculate the large order behavior of this
correlation function at   $k\rightarrow\infty$.
The main Lipatov's idea is to present the coefficients
$Z_k^{(n)}$ through a contour integral in the complex $g-$ plane:
\be
\label{15}
 Z_k\sim \int D\phi\oint\frac{dg}{g^{k+1}}e^{-S(\phi)},
\ee
where $S(\phi)$ is the action of the scalar field theory
and $D\phi$ is the standard measure for the functional integral which 
defines the theory (We confined ourself 
in this formula the case $n=0$ only. It corresponds to analysis
of the  Grand Partition Function
$Z(g)$ itself. The dependence on $n$ will be discussed a little bit
later.). If the theory possess the classical
instanton solution, then the calculation of the integral over $g$
can be done through steepest descent method.
 This method is
justified only for small $g$. But for  $k$ large, the 
integral over $g$ is dominated by the small $g$ contribution.
In particular,  for four dimensional $\phi^4$ field theory
the classical instanton solution has the property $\phi_{cl}\sim 1/\sqrt{g}$,
\cite{Lipatov}
and the classical action $S_{cl}\sim 1/g$ is parametrically
large. Thus, the semiclassical approximation is completely
justified.

In our case, when we consider 
the correlation function with external $\phi$
fields, and not the Grand Partition Function
$Z(g)$ itself,  we have to substitute   in the first approximation
the classical solution $\phi_{cl}$ in place of the 
external $\phi$ fields.
More precisely,
\be
\label{16}
 Z_k^{(n)}\sim \int D\phi\int\frac{dg}{g^{k+1}}
  \phi(x_1)\phi(x_2)...\phi(x_n)  e^{-S(\phi)}\sim 
\ee
\be
\int D\phi\int\frac{dg}{g^{k+1}}e^{-S(\phi_{cl})}
\phi_{cl}(x_1)\phi_{cl}(x_2)...\phi_{cl}(x_n) 
\sim \int_0^{\infty}
\frac{dg}{g^{k+1+\frac{n}{2}}}e^{-\frac{1}{g}}\sim(k+\frac{n}{2})!
\nonumber
\ee
In this formula
we took into account that the classical field depends on $g$ as 
$\phi_{cl}\sim 1/\sqrt{g}$ and the total number of external fields
in the correlation function
is equal $n$. It gives an additional factor $g^{\frac{n}{2}}$
in denominator. The  factorial dependence (\ref{16})
can be interpreted  as the rapid growing of the number
of Feynman graphs. In the spirit of the statistical approach \cite{JZJ}
 to evaluation of the Feynman graphs one could interpret the
same factorial dependence  as the large multiplicity of order $k$
  of the virtual states.

 The semiclassical approximation we have used in the 
derivation  (\ref{16})
is justified as far as number $(k+\frac{n}{2}) \gg 1$.
Only in this case the integral over $g$ is dominated by the small
$g$ contribution and instanton calculus can be applied.
It means that  this method is still
applicable   
 when  we calculate the dependence on the
number of external fields  $n\gg 1$. 
Thus,  by using this technique 
we can find out
 the dependence on number   $n$ which  is the main goal of the 
section.

After this reverence to scalar field theory we  
 come back to our  analysis of the appropriate correlation functions
  in QCD. 
 It can be shown \cite{Zhit3} that, up to some
unessential numerical factors, 
the analysis  of transverse moments can be reduced to the consideration
of the operators  
$\sim \bar{q}(\sigma_{\mu\nu}G_{\mu\nu})^nq$.
Thus, we arrive
to the following formula for the $n$-dependence 
of the transverse correlation function:
\be
\label{17}
\int DA\int\frac{dg^2}{(g^{2})^{k+1}}e^{-S(A_{cl})}
((G_{\mu\nu})_{cl})^n  \\   \nonumber
\sim \int_0^{\infty}
\frac{dg^2}{(g^2)^{k+\frac{n}{2}+1}}e^{-\frac{1}{g^2}}\sim(k+\frac{n}{2})!~,
\sim n!
 \ee
where we took into account that for the classical instanton solution
the field strength tensor is
$G_{\mu\nu}\sim 1/g$. The classical action for instanton is  
$S_{cl}\sim 1/g^2$. Besides that, we took into account 
that in order to provide the correct 
order in the perturbative expansion
   for $n$-th term  $\sim\alpha_s^n$,
  we need to take the $k=\frac{n}{2}$-th order  
 in the formula (\ref{17}).
It gives the appropriate
factor $(g^2)^\frac{n}{2}=g^n$
from the interaction of the external
gluons with themselves, see Fig.2.    Finally, we arrive
to the $n!$ dependence in the formula (\ref{17}), announced and 
intensively  employed earlier (\ref{10}).

 Few comments are in order. First of all,   the 
similar analysis of the longitudinal correlator (\ref{2})
does not lead to the $n!$ behavior found for the transverse
correlation function.
The technical reason for that is  
 the selection of the special kinematical structure
 $\sim(z_{\mu}q_{\mu})^n, z^2=0$ in (\ref{2}).
 An instanton, simply by kinematical reasons,
does not contribute to this  structure in the leading order.

Now  we wish to understand  the difference, mentioned above, 
between longitudinal and transverse correlators
in terms of standard Feynman diagrams.
As we discussed earlier, the factorial dependence
in the  field theory is the simple reflection 
of the rapid growing of the Feynman graphs. 
Thus, we have to be surprised not by the fact that 
such factorial dependence appeared, but rather,
the fact that sometimes it does not show up.
Therefore, let us try to understand why
the extracting of the longitudinal structure 
suppresses the most of the graphs?

To answer on this question, let us consider 
a typical graph with three-gluon vertex, Fig.3 (exactly
such kind of graphs, at least in the limit of large 
number of colors,
 gives the most important
statistical contribution) and take the longitudinal projection
for the external gluons $\sim (A_{\mu}z_{\mu})^2$.
We wish to extract the structure $\sim (q_{\mu}z_{\mu})^2$
after completing the calculation. The direct computations
demonstrate that the corresponding contribution to this structure is zero.
The only nonzero result comes from the graphs like
Fig.4, where the external gluons interact with quarks,
but not with themselves. It is clear, that the number of
such diagrams does not demonstrate the factorial growth.
  
 The explanation is simple:   transverse and longitudinal 
structures are quite different because   
   transverse
gluons (in contrast to longitudinal ones) are physical degrees
of freedom and can not be rotated away by a gauge transformation.
Both methods presented above
(instanton calculus and diagrammatic approach) lead
to the same conclusion and agree with each other.
  \vskip .3cm
\noindent
{\bf 4.Nonlocal condensate and constituent mass of the quark.}
  \vskip .3cm
The nonlocal condensate (NLC) 
   has been introduced into the theory (to my knowledge)
 within operator product expansion (OPE) 
technique by Shuryak more than 10 years ago
\cite{Shuryak}.\footnote{The term NLC itself appeared later,
\cite{Raduyshkin}.}
This object   naturally arises
in the analysis of the heavy-light quark system, $\bar{q}Q$.
Indeed, if we consider along with \cite{Shuryak},\cite{Rad2}
 the correlation function
$\la T\{ \bar{q}Q(x),\bar{Q}q(0)\}\ra$, describing this system,
we end up (in the limit $M_Q\rightarrow\infty$) with the object
which is completely factorized from the heavy quark and which 
is called   NLC:
\be
\label{Q} 
\la T\{ \bar{q}Q(x),\bar{Q}q(0)\}\ra
\sim  \la\bar{q}(x)P\exp(ig\int_0^xA_{\mu}dx_{\mu})q(0)\ra
+ perturb. ~ part .
\ee
All nontrivial, large distance physics of
the system is hidden there. The NLC is
gauge invariant, well defined object and 
thus, it   deserves to be studied in a more detail.
As we mentioned,  this system may provide a
definition of the constituent quark in QCD.
We shall discuss this important point a little bit later, but now we
would like to note that 
 for the large Euclidean $x$ the NLC is dominated
by the lowest state contribution and thus \cite{Rad2}:
\be
\label{18}
   \la\bar{q}(x)P\exp(ig\int_0^xA_{\mu}dx_{\mu})q(0)\ra
\stackrel{-x^2\rightarrow\infty}{\sim}\exp(-\bar{\Lambda}\sqrt{-x^2}),
\ee
where $\bar{\Lambda}$ is the lowest energy level
of the mesons in the heavy quark effective theory (HQET)
(see, e.g. review \cite{Wise}).
 
Another interesting application of the NLC is the analysis
of the hadron properties within QCD sum rules \cite{Raduyshkin},\cite{Rad3}.
However,
we shall not discuss any applications 
of the NLC in this letter. Instead we concentrate on the analysis
of the NLC itself and its general properties.

It is convenient to parametrize the coordinate dependence
of the NLC by a Laplace-type representation\cite{Raduyshkin}:
\be
\label{19}
  \la\bar{q}(x)P\exp(ig\int_0^xA_{\mu}dx_{\mu})q(0)\ra\equiv
\la\bar{q}q\ra S(x^2),~~
S(x^2)=\int^{\infty}_{0}e^{\frac{sx^2}{4}}f(s)ds.
\ee
From its definition it is clear that the $n-$ th moment
of the spectral function $f(s)$ can be expressed 
in terms of local operators with $2n$ derivatives:
\be
\label{20}
 \frac{1}{n!}(\frac{x^2}{4})^n\la\bar{q}q\ra\int^{\infty}_{0}f(s)s^nds=
 \frac{1}{(2n)!}\la\bar{q}(x_{\mu}D_{\mu})^{2n} q\ra  
  \ee
 Thus, the entire information about NLC is hidden now 
in the properties of the spectral function $f(s)$
or (which is the same) is coded by the behavior of the high
dimensional condensates $\sim\la\bar{q}(D_{\mu})^{2n} q\ra$.

What do we know about properties of the $f(s)$?
 First of all, we know its
normalization and the first moment:
\be
\label{21}
\int^{\infty}_{0}f(s)ds=1,  \nonumber
\ee
\be 
\int^{\infty}_{0}f(s)sds=
\frac{\la\bar{q}D_{\mu}D_{\mu}q\ra}{2\la\bar{q}q\ra}=
\frac{\la\bar{q}ig\sigma_{\mu\nu}G_{\mu\nu} q\ra}{4\la\bar{q}q\ra}=
\frac{m_0^2}{4}\simeq 0.2 GeV^2 , 
\ee
where we introduced the standard notation (see e.g.\cite{Shif2})
 $m_0^2\simeq 0.8 GeV^2$ for the mixed 
quark-gluon vacuum condensate of the dimension five.

Besides that, we know the asymptotic of the NLC 
for large euclidean $x$ (\ref{18}).
The transformation (\ref{19}) converts  this information
into the fixing of the asymptotic behavior of the spectral
function at small $s$ : $f(s\rightarrow 0)
\sim  \exp(-\frac{\bar{\Lambda}^2}{s})$.

In addition to these, well established results,
we have some more constraints. First, we know the next,
   high dimensional condensate, discussed earlier
\cite{Zhit3},\cite{Zhit5}. Second,   
 the  large $n$- behavior for such condensates 
  can  be found   by the method elaborated in the previous section.

Let us start from the analysis of the next moment of the
spectral function $f(s)$.  The   formula (\ref{20})
gives the following relation between the second moment of $f(s)$ and 
the vacuum condensate of dimension seven:
\be
\label{22}
 \frac{3x^4}{4}\la\bar{q}q\ra\int^{\infty}_{0}f(s)s^2ds=
 \la\bar{q}(x_{\mu}D_{\mu})^4 q\ra =\\     \nonumber
x^4\{\frac{-3}{96} \la\bar{q}g^2\sigma_{\mu\nu}G_{\mu\nu}
 \sigma_{\lambda\sigma}G_{\lambda\sigma}q\ra
+ \frac{3}{48} \la\bar{q}g^2G_{\mu\nu}G_{\mu\nu}q\ra\},
 \ee
where we evaluated the condensate 
$\la\bar{q} D_{\mu}D_{\nu}D_{\lambda}D_{\sigma} q\ra$
in a standard way, by reducing it to the operator which depends
only on $G_{\mu\nu}$.
\cite{Zhit3},\cite{Zhit5}.
To estimate the right hand side of the
eq.(\ref{22}), we introduce the coefficient of   
``nonfactorizability" $K$ for the condensates of dimension seven 
(one and the same for all operators). In particular,
$$ \la\bar{q}g^2\sigma_{\mu\nu}G_{\mu\nu}
 \sigma_{\lambda\sigma}G_{\lambda\sigma}q\ra
= \frac{-K}{3}\la\bar{q}q\ra \la g^2G_{\mu\nu}^aG_{\mu\nu}^a\ra , $$
$$\la\bar{q}g^2G_{\mu\nu}G_{\mu\nu}q\ra=
 \frac{K}{6}\la\bar{q}q\ra \la g^2G_{\mu\nu}^aG_{\mu\nu}^a\ra . $$
(In this formula  $K=1$  if the factorization prescription would work).
 The coefficient $K$ was estimated numerically in
\cite{Zhit3},\cite{Zhit5} and was found to be  from $3$ to $4$.
 There is an additional,
independent estimate which supports this value  for $K$ and which
 will be presented at the end of the section.

Finally, we arrive to the following value for the
second moment of the spectral function:
\be
\label{23}
\int^{\infty}_{0}f(s)s^2ds=
 \frac{K}{36}\la g^2G_{\mu\nu}^aG_{\mu\nu}^a  \ra
\simeq 0.04 \div 0.05 GeV^4,
\ee
where we use the standard value for the gluon condensate
$\la\frac{\alpha_s}{\pi}G_{\mu\nu}^2\ra\simeq 1.2 \cdot 10^{-2}GeV^4$,
\cite{Shif2} and 
the estimated value for the coefficient $K=3\div 4 .$

Our next constraint comes from the analysis of the 
asymptotically distant terms in the formula (\ref{20}).
 We use PCAC in order to relate the matrix elements and vacuum
condensates:  
 \be
\label{24}
 \la 0|\bar{d}i\gamma_5(x_{\mu}D_{\mu})^{2n}u|\pi\ra=\frac{-2}{f_{\pi}}
\la 0|\bar{q}(x_{\mu}D_{\mu})^{2n} q |0\ra,  
  \ee
By repeating the previous arguments
we arrive to the following conclusion (see footnote 2  on page 4):
\be
\label{26}
 \la  \bar{q}(x_{\mu}D_{\mu})^{2n} q  \ra\sim
 (x^2)^n   \la  \bar{q}( D^2)^{n} q  \ra\sim
 (x^2)^n   \la  \bar{q}( ig\sigma_{\mu\nu}G_{\mu\nu})^{n} q  \ra
\sim  n!
  \ee
This result combined with the definition (\ref{20}) gives the
following constraint on the asymptotic behavior
of the spectral function $f(s)$:
\be
\label{27}
\int^{\infty}_{0}s^nf(s)ds \sim 1.
 \ee
 Let us note the absence of the factorial dependence on the right hand side of 
(\ref{27}). 
It means that $f(s)$ falls off faster than any exponential function.
We parametrize such a dependence at large $s$ by step fuction:
  $f(s )\sim \Theta (\sigma^2 -s)$, where $\Theta(x)-$ function
is defined in the standard way: $\Theta(x> 0)=1, ~\Theta(x< 0)=0$.

Combining this result with the small $s$ behavior in the simplest way,
we arrive at the  following Ansatz for the spectral density $f(s)$: 
\be
\label{28}
f(s)=N\cdot e^{-\frac{\Lambda^2}{s}}\Theta(\sigma^2-s) 
\ee
with three dimensional parameters: $N,\Lambda, \sigma$.

 The Fourier transformed  function $\tilde{S(p)}$  
 with Ansatz (\ref{28}) takes  the following form:
\be
\label{31}
\tilde{S(p)}\sim   \int^{\sigma^2}_{0}\frac{ds}{s^2}
e^{\frac{p^2-\Lambda^2}{s}} \sim
\frac{1}{\Lambda^2-p^2}\cdot e^{\frac{-\Lambda^2+p^2}{\sigma^2}}. 
\ee
Few comments are in order.
 First of all, 
all formulae in this section, starting from
(\ref{18}), are derived in the Euclidean space
(imaginary time), $x^2 <0,~p^2 <0$. However we expect that the obtained
analytical properties  ({\bf pole}, in particular) will 
not be changed after the continuation to  
Minkowski space is made.
In this sense the formula (\ref{31}) demonstrates that we have dealt with a
{\bf ``constituent quark" with mass $\Lambda$} and width $\sigma$. 
This determines the physical meaning of our phenomenological parameters
introduced by Ansatz (\ref{28})\footnote{We expect,
of course, that the constituent quark mass depends on momentum $p^2$.
We believe that this dependence  
  would come into the game 
if the corrections  $\sim 1/n$ in eq.(\ref{27})would be taken into account.
  In particular, more complex Ansatz for the spectral
density $f(s)\rightarrow f(s)\sum_{k=0}a_ks^k, ~a_0\neq 0$
does not change the asymptotical behavior, but  leads
to the momentum dependence of the constituent quark mass.
Unfortunately, we can not reconstruct  
these, non-leading terms, by the described method.}. 

  Our next comment is the qualitative remark that our Ansatz (\ref{28})
for the spectral density $f(s)$ describes the well-localized function
 (at $s_0\simeq\sigma^2$) almost 
for arbitrary parameters $\Lambda, \sigma$. Thus, 
for any reasonable choice of these parameters
we would expect that the following feature of the Ansatz (\ref{28})
is fulfilled:
\be
\label{32}
\frac{\int^{\infty}_{0}s^2f(s)ds}{(\int^{\infty}_{0}sf(s)ds)^2} \simeq
\frac {s_0^2}{(s_0)^2}\simeq 1.
\ee
 It should be considered as an independent check of 
  the nonfactorizability of the mixed vacuum condensate 
of dimension seven with factor $K\simeq 3\div 4$.  

We are prepared   for some numerical estimates now. The $\sigma$ 
parameter is fixed   by the mixed vacuum condensate (\ref{21})
and is equal to $\sigma^2\simeq 0.3 GeV^2$.  The parameter $\Lambda$  
  presumably should coincide with the constituent quark mass, i.e.
$\Lambda\simeq 0.35 GeV$.  
We sketch  the NLC on Fig.5 where we use    
these numerical parameters.

Our last remark is a very general observation that the 
{\bf origin} for the Gaussian
behavior of the light cone $wf$ (\ref{!})
(it corresponds  to   the harmonic oscillator $wf$ in the
quark model, see(\ref{qm}))
and  the very specific behavior  
(\ref{27},\ref{28}) of the vacuum nonlocal condensate
  {\bf is one and the same}:
It is the factorial dependence  of the pion matrix elements (\ref{26})
in the former case and the same factorial dependence
of the vacuum condensates $\la \bar{q}(D^2)^nq\ra\sim n!$ in the latter one. 
It determines both these phenomena.
     \vskip .3cm
\noindent
{\bf 5.Conclusion.}
  \vskip .3cm

Let me formulate the main results of this letter.
  
$\bullet$ We formulated some constraints on the nonperturbative 
light cone $wf$. The analysis was based on the very general
methods like dispersion relation and duality, and thus,
the obtained results
should be considered as direct consequence of QCD.
 We found the Gaussian asymptotics for the $wf$,
which we believe, not accidentally coincides with the harmonic 
oscillator $wf$ from valence quark model.

$\bullet$ We reconstructed the NLC, the fundamental object
of QCD, by analysing its few moments and the
asymptotical behavior.
  
We hope that this letter can motivate
the future investigations in the following direction:    
How one can  understand  the valence quark model
(which    is the main tool in the hadronic
physics for already three decades) in terms of
QCD Lagrangian? .

 \noindent
{\bf \bf 6. Acknowledgements.}
  \vskip .3cm
 
It is a pleasure to thank Stan Brodsky for   encouragement
and many interesting discussions related to the subject of this paper.
I am thankful to A.Radyushkin for the critical remarks.
 I wish to thank the organizers and participants of the 
4-th Light Cone Meeting, Seattle 1994, where some results
 of this paper were presented.

This work is supported by the Texas National Research 
Laboratory Commission under  grant \# RCFY 93-229.
\vskip .3cm
 
\noindent
{\bf Figure Captions.}
\vskip .3cm
Fig.1 The asymptotic behavior of the two -point correlation function
at one loop level.

Fig.2 The large order corrections to the  correlation function
which lead to $n!$ behavior. 

Fig.3 A typical graph with three-gluon vertex which contributes
to the transverse, but not longitudinal structure. 

Fig.4
A typical graph which contributes to the longitudinal structure. 

Fig.5 The NLC
   $\frac{\la\bar{q}(x)P\exp(ig\int_0^xA_{\mu}dx_{\mu})q(0)\ra }
{\la\bar{q}q\ra}= S(x)$ as function of distance for parameter
$\Lambda\simeq 0.35 GeV$.

\end{document}